# Substrate-Modulated Reductive Graphene Functionalization

Ricarda A. Schäfer[a], Konstantin Weber[b], Frank Hauke[a], Vojislav Krstic[c], Bernd Meyer[b], and Andreas Hirsch[a]*

[a] Ricarda A. Schäfer, Dr. Frank Hauke, Prof. Dr. Andreas Hirsch
Department of Chemistry and Pharmacy & Joint Institute of Advanced Materials and Processes (ZMP)
Friedrich-Alexander-Universität Erlangen-Nürnberg
Henkestr. 42, 91054 Erlangen, Germany
E-mail: andreas.hirsch@fau.de

[b] Konstantin Weber, Prof. Dr. Bernd Meyer
Interdisciplinary Center for Molecular Materials (ICMM) and Computer-Chemistry-Center (CCC),
Friedrich-Alexander-Universität Erlangen-Nürnberg, Nägelsbachstraße 25, 91052 Erlangen, Germany

[c] Prof. Dr. Vojislav Krstic
Department of Physics, Friedrich-Alexander-Universität Erlangen-Nürnberg, Erwin-Rommel-Straße 1, 91058 Erlangen, Germany

**Abstract:** Covalently functionalizing mechanical exfoliated monolayer and bilayer graphenides with λ-iodanes led to the discovery that the monolayers supported on a $SiO_2$ substrate are considerably more reactive than bilayers as demonstrated by statistical Raman spectroscopy/microscopy. Supported by DFT calculations we show that ditopic addend binding leads to much more stable products than the corresponding monotopic reactions due to much lower lattice strain of the reactions products. The chemical nature of the substrate (graphene *versus* $SiO_2$) plays a crucial role.

The systematic exploration of the general principles of graphene functionalization will allow the tuning of the pro-cessibility and the properties, which is the key for the devel-opment of technological applications of this 2D nano material. In this regard, the investigation and establishment of functionalization sequences on graphene deposited on substrates constitutes a very attractive scenario, since in contrast to bulk chemistry well defined conditions for the chemical transformations themselves and a very straightforward product analysis are provided.[1] In previous studies it has been demonstrated that on-surface modifications of graphene depend on the electronic,[2-3] topological,[4] and chemical structure of the substrate. In this study we consider three possible reaction scenarios for the chemical nature of the substrate surface (surface functionalities): a) the substrate surface is inert and only one-sided (monotopic) additions are possible[5-7] (Figure 1a), b) the substrate contains surface functionalities that after the initial attack of the addend R can undergo a subsequent ditopic addition to the graphene[8] (Figure 1b), and c) the substrate is shielded by a second graphene layer. In this case the question arises whether after an initial attack of an addend R to the outer graphene layer a subsequent covalent bond formation with the graphene layer underneath is a preferred process or not (Figure 1c).[9]

In order to investigate these three reaction scenarios de-picted in Figure 1, we choose the reductive functionalization that we have recently introduced for the functionalization of both bulk graphene and surface deposited graphene.[10-14] In contrast to other routes, here, the graphene is negatively charged prior to the subsequent chemical functionalization. These activated layers, known as graphenides,[15] are then covalently attacked by suitable electrophiles such as alkyl iodides, diazonium salts, and λ -iodane compounds.[11-13, 16]

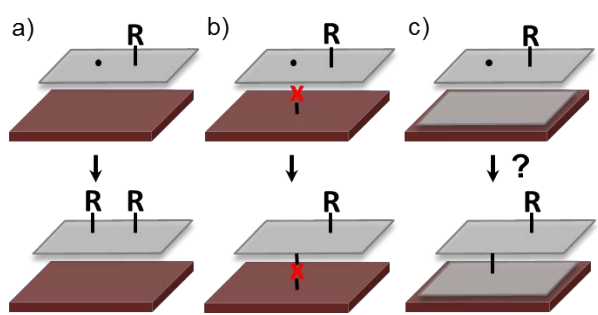

**Figure 1**. Possible reaction scenarios for the functionalization of a graphene monolayer on a substrate considered in this study: a) monotopic reaction sequence on an inert substrate, b) ditopic reaction sequence involving a second step binding of a reactive surface functionality X, and c) potential ditopic reaction sequence involving a C-C bond formation with a second graphene layer underneath covering the substrate.

Due to the reductive activation, the Fermi level is shifted and the reactivity of the graphene layer towards radical reactions is increased. In general, radical reactions on graphene are accompanied by the formation of a dangling bond. In the case of a monolayer functionalization on a surface, it can be anticipated that the substrate, providing functional entities, can interact with this dangling bond of the deposited graphene flake, changing a monotopic addition sequence into the ditopic functionalization scenario b). Bilayer graphene, however, provides an ideal model architecture for the scenarios a) or c) (Figure 1). In this study we directly and simultaneously compare the reactivity of reduced mono- and bilayer graphene deposited on a $SiO_2$ surface towards the reaction with *bis*-(4-(*tert*-butyl)phenyl)iodonium hexafluorophosphate (see ESI, Figure S1), which allows us for the first time to address the question raised in Figure 1. A detailed Raman spectroscopic and microscopic analysis of the reaction products was corroborated by a systematic density-functional theory (DFT) study.

With the purpose to address the question of a monotopic versus a ditopic functionalization scenario and to investigate the different reactivity of mono- and bilayer graphene, we mechanically exfoliated graphene and obtained one distinct flake which exhibits a monolayer area adjacent to a bilayer region. This flake was reductively activated by the addition of a blue solution of solvated $NaK_3$ in 1,2-dimethoxyethane (DME) (**G$_A$**, Scheme 1) – experimental conditions see ESI.[10, 12, 14, 17] The activation was followed by the addition of *bis*-(4-(*tert*-

butyl)phenyl)iodonium hexafluorophosphate in DME causing an arylation of the deposited graphene sheet (**G$_B$**, Scheme 1). This arylation is based on the formation of an aryl radical, which is formed due to a charge transfer reaction from the charged graphene to the iodonium salt.[13, 18]

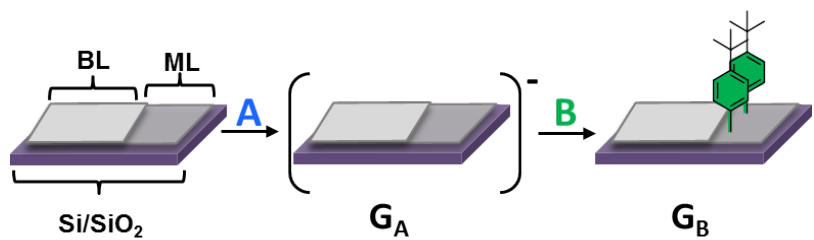

**Scheme 1.** Arylation of a graphene flake (monolayer (ML) and bilayer (BL) graphene) deposited on a Si/SiO$_2$ substrate. A) NaK$_3$ in DME and B) *bis*-(4-(*tert*-butyl)phenyl)iodonium hexafluorophosphate in DME.

In order to investigate the reaction product, we applied statistical Raman spectroscopy (SRS) and microscopy (SRM)[12] before and after the reductive activation as well as after addition of the trapping reagent. The mapping of the three prominent Raman bands of graphene, namely the D-band at 1350 cm$^{-1}$, the G-band at 1582 cm$^{-1}$, and the 2D-band at 2700 cm$^{-1}$ and their correlation, clearly allows for the identification of the mono- and bilayer areas present in the mechanical exfoliated flake (Figure 2a,b). In the $I_{2D}/I_G$ map, the monolayer area appears yellow to red, which correlates to a $I_{2D}/I_G$ ratio of 3.0 to 4.0. Point spectra of this area A show a Lorentzian-shaped 2D band with a FHWM of 27 cm$^{-1}$ (Figure 2g), clearly demonstrating its single layer character.[19] In contrast, the blue and green colored parts of the map in area B display a $I_{2D}/I_G$ ratio of 0.8 to 1.2 and a broad 2D band. The $I_D/I_G$ ratio map of the same flake, which can be used for the quantification of defects,[20] illustrates the flawless carbon lattice of the starting material. After the functionalization the image has changed. Now the $I_D/I_G$ ratio map displays a light blue region with a $I_D/I_G$ ratio of 0.6 and a green region with a $I_D/I_G$ ratio of 2.0, which can be associated with the bilayer part B and the monolayer part A of the graphene flake (Figure 2d). Based on the spectral information, the mean defect distance $L_D$ can be calculated.[12] The respective $L_D$ map (Figure 2e) nicely confirms the results obtained from the $I_D/I_G$ ratio analysis, as the $L_D$ of the area A, with a maximum intensity value of 8, is drastically lower than the values obtained for area B. The following translation of the defect distance $L_D$ into the degree of functionalization $\Phi$ exhibits values of 0.05 to 0.06 for the monolayer section A, in comparison to $\Phi$-values of 0.02 obtained for the bilayer region B (Figure 2f). Thus, it becomes apparent that in the monolayer region A a higher amount of addends have been bound to the basal plane, in contrast to the bilayer section B where almost no functionalization has occurred, which is also nicely corroborated by the respective single point spectra presented in Figure 2g and 2h.

In order to ensure that functionalization is only due to the depicted covalent binding of the electrophile or surface functionalities of the substrate, we also carried out an in situ characterization (under strict inert gas conditions) of an activated mono/bilayer graphene flake by means of SRM. The corresponding $I_D/I_G$ ratio map (Figure S3) showed clearly that no defects are introduced by side reactions with the solvent since the $I_D/I_G$ ratio of 0.2 for the monolayer area and a $I_D/I_G$ ratio of 0.15 for the bilayer section stayed constant. Furthermore, a shift of the G-band position as well as a decrease in the 2D-band intensity has been observed. Both are known to be characteristic for doping effects.[21] Therefore, we can safely exclude any covalent side reactions.

For a better understanding of the nature of the functionalization in the case of graphene deposited on a substrate and for a comparison of the results obtained in the bulk functionalization scenario, we also applied temperature-dependent Raman spectroscopy on our samples (Figure S4).

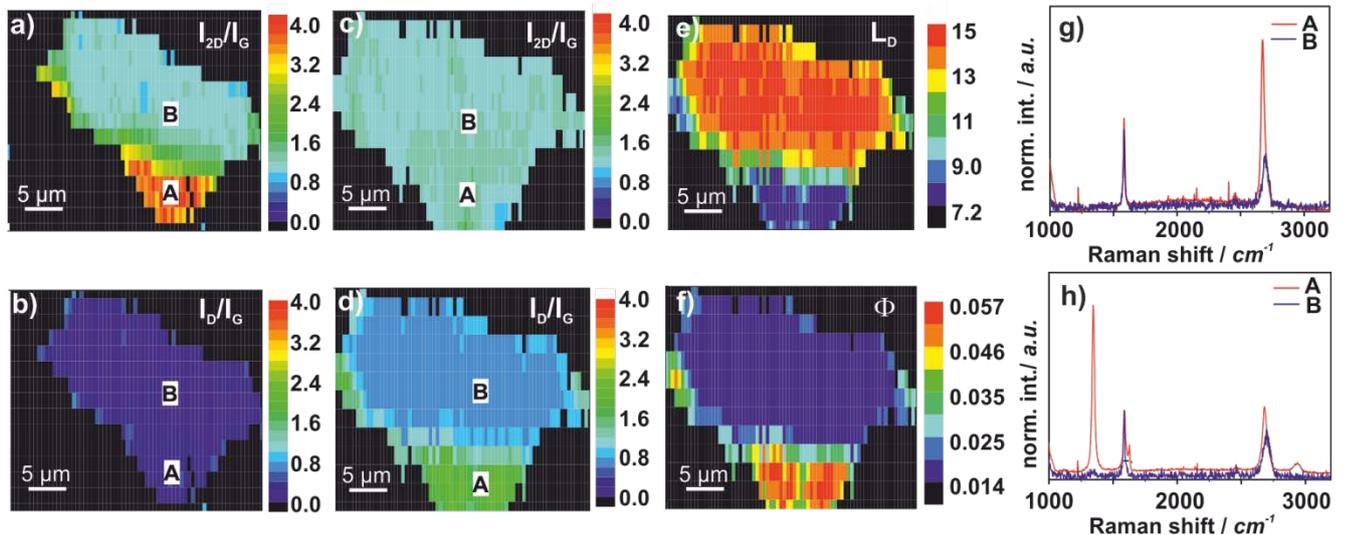

**Figure 2.** Scanning Raman microscopy images of an arylated graphene flake (monolayer region A, bilayer region B): a) $I_{2D}/I_G$ and b) $I_D/I_G$ map of pristine mechanically exfoliated graphene, c) $I_{2D}/I_G$ and d) $I_D/I_G$ map of functionalized graphene, e) $L_D$ map of functionalized graphene, f) $Φ$-map of functionalized graphene, g) Raman spectra of monolayer area A and bilayer area B of pristine graphene and h) Raman spectra of monolayer area A and bilayer area B of functionalized graphene.

Starting from 40 °C, the functionalized graphene flake was mapped in intervals of 50 °C. The SRM images of the $I_D/I_G$ ratio exhibit a constant value between 0 and 0.4 for the bilayer area. The small D-band vanishes at 100 °C. In the monolayer section of the flake at this temperature a very prominent D-band, indicative for a high degree of covalent functionalization, is still observed. The $I_D/I_G$ values (3.0 to 3.5) in the monolayer area at 40 °C decrease to values of 1.6 to 2.0 at 150 °C and finally to 0.6 at 300 °C. The evolution of the depicted point spectra nicely visualizes the trend obtained for the whole map. In addition, this temperature-dependent

$I_D/I_G$ profile perfectly correlates with the thermogravimetric data of a bulk sample (Figure S5), again proving the successfully covalent binding of the aryl addends.

Based on all these experimental results it becomes evident that the reductive functionalization of mono- and bilayer graphene leads to a highly preferred addend binding of the monolayers directly deposited on the Si/SiO$_2$ substrate. But why is that the case? As mentioned above, we anticipate that the chemical nature of the substrate should play an important role and could give rise to secondary attacks *antaratopic* to the addends already bound. Hence, the higher degree of functionalization of the monolayer area can be attributed to scenario b), where a reactive structure delivers a possible reactant for a ditopic addition involving reactive surface functionalities of the substrate. In our case, the substrate surface contains SiOH groups and residual water.[3] On the other hand, in the case of the graphene bilayer structure of area B, the bottom graphene layer acts as an inert/unreactive support and resembles scenario a).

For a deeper understanding of the reactivity patterns we performed theoretical calculations on the DFT level (for details see ESI). For this purpose, the binding energy of the addends was calculated by taking the energy difference between functionalized and unfunctionalized graphene mono- and bilayers and the radical species. For single addends the binding energy represents the strength of the C–C bond which is formed upon covalent bond formation. To simplify the calculations, a methyl instead of the 4-(*tert*-butyl)phenyl group was used. Since the methyl group is the smallest organic unit to form C–C bonds with graphene, a lower limit for the proximity effects will be obtained when the degree of functionalization is increased in the calculations.

For a single methyl group on a freestanding graphene monolayer (Figure 3b) we find a rather weak C–C bond strength of only 48 kJ/mol. This is considerably less than what we obtained for the C–C bond strength of methyl groups on fullerenes (155 kJ/mol) and on carbon nanotubes (179 and 100 kJ/mol for (6,0) and (10,0) tubes, respectively). Such a weak bond strength should lead to a rather low degree of monotopic (one-sided) functionalization of graphene and corresponds to scenario a). However, if the bottom of the graphene layer is simultaneously saturated at a neighboring C atom as in scenario b), for example by the adsorption of an OH group (Figure 3a), the C–C bond strength of the methyl group increases to 209 kJ/mol. Thus, the experimentally observed functionalization of monolayer graphene on Si/SiO$_2$ is most likely accompanied by a significant addition of residual species from the Si/SiO$_2$ substrate, enabling a strain free ditopic reaction pathway as shown in scenario b). In which particular way the bottom side of the graphene layer is attacked by the surface, be it a SiOH-group or another functionality (X, Figure 1b), plays a minor role. For systematic calculations we used a Ni(111) surface as a simple tool to guarantee that all C atoms that need to be saturated automatically have a binding partner.[8] Since the Ni(111) surface has the same lattice constant as graphene, all C atoms of one graphene sublattice sit on-top of Ni and can form a bond to an Ni atom whenever they need to be saturated. Indeed, for the graphene monolayer on the Ni(111) substrate (Figure 3c), the C–C bond strength with 194 kJ/mol remains almost the same as for the OH-saturated freestanding layer.

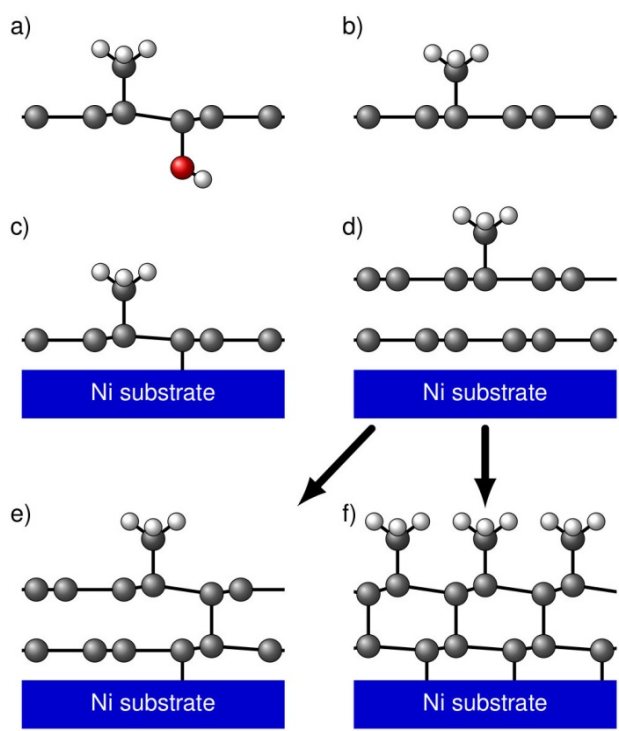

**Figure 3.** Models of functionalized mono- and bilayer graphene. a) Freestanding ditopic functionalized monolayer graphene, b) freestanding monotopic functionalized mono-layer graphene, c) functionalized monolayer graphene on Ni substrate, d) functionalized bilayer graphene on Ni substrate, e) and f) possible interconnections of functionalized bilayer graphene on Ni substrate.

For the supported bilayer we obtain a C–C bond strength for the methyl group of 60 kJ/mol (Figure 3d). This is basically identical to the result for monotopic functionalization of freestanding monolayer graphene, which is exactly what is expected if it is assumed that the second graphene sheet in the bilayer shields the upper layer from being saturated at the bottom as shown in scenario a). However, also for the bilayer a saturation mechanism can, in principle, be anticipated: two C atoms with on-top stacking in the bilayer form a bond and, in turn, a neighboring C atom in the second graphene layer undergoes covalent binding to the surface (Figure 3e) as proposed in scenario c). In our geometry optimizations, however, such a scenario turned out to be unstable. The reason is the high cost in strain energy since the C–C distance in the bilayer of 3.26 Å has to reduce to typical values for a C–C bond of about 1.4–1.6 Å. Although the sp$^3$ hybridization of the functionalized C atom in the upper layer and the saturated C atom in the lower layer help to bring the two C atoms with on-top stacking closer together, it is not sufficient that a single C–C bond can form. This nicely explains the low degree of experimental functionalization for the bilayer graphene part.

Theoretically, this limitation might be overcome at higher degrees of bilayer functionalization, when several interlayer C–C bonds could mutually stabilize each other. Indeed, for a functionalization degree of 50 % or 100 %, stable structures with interlayer C–C bonds are found in the geometry optimization (Figure 3f). However, the binding energy of the methyl groups is dramatically decreased to even negative values, i.e., the functionalized bilayer would gain energy by spontaneous repulsion of methyl groups. At a functionalization degree of 50 % or 100 %, the methyl groups are much too densely packed and repel each other. At a functionalization degree of 33 %, the highest possible coverage with no methyl groups on neighboring carbon sites (within the same sublattice), the interlayer C–C bonds are already unstable and do not form. This is reflected by a methyl binding energy of 49 kJ/mol, which is already close to the value of 60 kJ/mol for the dilute limit of bilayer functionalization.

In conclusion, we have discovered that the reductive functionalization of monolayer graphene on a substrate leads to a higher degree of functionalization than the corresponding functionalization of bilayer graphene. Significantly, the chemical reactivity of the graphene monolayer supporting substrate plays a critical role. A reactive substrate can stabilize intial adducts by saturating dangling bonds *antaratopic* to external addends. This pathway is absent if the substrate is inert, as exemplified in bilayer graphene involving a graphene buffer layer. Our experimental results are nicely corroborated by theoretical DFT calculations. If exclusively *supratopic* additions are possible, a very high amount of strain energy on the C-lattice accumulates continuously. This limits the maximum amount of addend binding considerably. As a consequence, much lower degrees of functionalization can be expected compared with ditopic reactions allowing also for *antaratopic* addend binding.

**Acknowledgements**

The authors thank the Deutsche Forschungsgemeinschaft (DFG) for financial support through Collaborative Research Center SFB 953 "Synthetic Carbon Allotropes" (Project A1, B12, and C1). The research leading to these results has re-ceived partial funding from the European Union Seventh Framework Programme under grant agreement no.604391 Graphene Flagship. K.W. thanks the Fonds der Chemischen Industrie FCI for a Chemiefonds Fellowship.

**References**

[1] A. Criado, M. Melchionna, S. Marchesan, M. Prato, *Angew. Chem. Int. Ed.* **2015**, *54*, 10734-10750.
[2] Q. H. Wang, Z. Jin, K. K. Kim, A. J. Hilmer, G. L. C. Paulus, C.-J. Shih, M.-H. Ham, J. D. Sanchez-Yamagishi, K. Watanabe, T. Taniguchi, J. Kong, P. Jarillo-Herrero, M. S. Strano, *Nat. Chem.* **2012**, *4*, 724-732.
[3] X. Fan, R. Nouchi, K. Tanigaki, *J. Phys. Chem. C* **2011**, *115*, 12960-12964.

[4] Q. Wu, Y. Wu, Y. Hao, J. Geng, M. Charlton, S. Chen, Y. Ren, H. Ji, H. Li, D. W. Boukhvalov, R. D. Piner, C. W. Bielawski, R. S. Ruoff, *Chem. Commun.* **2013**, *49*, 677-679.
[5] R. Sharma, J. H. Baik, C. J. Perera, M. S. Strano, *Nano Lett.* **2010**, *10*, 398-405.
[6] Q. H. Wang, C.-J. Shih, G. L. C. Paulus, M. S. Strano, *J. Am. Chem. Soc.* **2013**, *135*, 18866-18875.
[7] G. L. C. Paulus, Q. H. Wang, M. S. Strano, *Acc. Chem. Res.* **2013**, *46*, 160-170.
[8] W. Zhao, J. Gebhardt, F. Späth, K. Gotterbarm, C. Gleichweit, H.-P. Steinrück, A. Görling, C. Papp, *Chem. Eur. J.* **2015**, *21*, 3347-3358.
[9] L. Yuan, Z. Li, J. Yang, J. G. Hou, *PCCP* **2012**, *14*, 8179-8184.
[10] J. M. Englert, C. Dotzer, G. Yang, M. Schmid, C. Papp, J. M. Gottfried, H.-P. Steinrück, E. Spiecker, F. Hauke, A. Hirsch, *Nat. Chem.* **2011**, *3*, 279-286.
[11] A. Hirsch, J. M. Englert, F. Hauke, *Acc. Chem. Res.* **2013**, *46*, 87-96.
[12] J. M. Englert, P. Vecera, K. C. Knirsch, R. A. Schäfer, F. Hauke, A. Hirsch, *ACS Nano* **2013**, *7*, 5472-5482.
[13] F. Hof, R. A. Schäfer, C. Weiss, F. Hauke, A. Hirsch, *Chem. Eur. J.* **2014**, *20*, 16644-16651.
[14] K. C. Knirsch, R. A. Schäfer, F. Hauke, A. Hirsch, *Angew. Chem. Int. Ed.* **2016**, *55*, 5861-5864.
[15] A. Pénicaud, C. Drummond, *Acc. Chem. Res.* **2013**, *46*, 129-137.
[16] J. M. Englert, K. C. Knirsch, C. Dotzer, B. Butz, F. Hauke, E. Spiecker, A. Hirsch, *Chem. Commun.* **2012**, *48*, 5025-5027.
[17] J. L. Dye, *Angew. Chem. Int. Ed.* **1979**, *18*, 587-598.
[18] C. K. Chan, T. E. Beechem, T. Ohta, M. T. Brumbach, D. R. Wheeler, K. J. Stevenson, *J. Phys. Chem. C* **2013**, *117*, 12038-12044.
[19] L. M. Malard, M. A. Pimenta, G. Dresselhaus, M. S. Dresselhaus, *Physics Reports* **2009**, *473*, 51-87.
[20] L. G. Cançado, A. Jorio, E. H. M. Ferreira, F. Stavale, C. A. Achete, R. B. Capaz, M. V. O. Moutinho, A. Lombardo, T. S. Kulmala, A. C. Ferrari, *Nano Lett.* **2011**, *11*, 3190-3196.
[21] A. Das, S. Pisana, B. Chakraborty, S. Piscanec, S. K. Saha, U. V. Waghmare, K. S. Novoselov, H. R. Krishnamurthy, A. K. Geim, A. C. Ferrari, A. K. Sood, *Nat. Nanotechnol.* **2008**, *3*, 210-215.

# Substrate-Modulated Reductive Graphene Functionalization

*Ricarda A. Schäfer, Konstantin Weber, Frank Hauke, Vojislav Krstic,*
*Bernd Meyer and Andreas Hirsch\**

Department of Chemistry and Pharmacy and Joint Institute of Advanced Materials and Processes (ZMP), Friedrich-Alexander-Universität Erlangen-Nürnberg, Henkestraße 42, 91054 Erlangen, Germany

Interdisciplinary Center for Molecular Materials (ICMM) and Computer-Chemistry-Center (CCC), Friedrich-Alexander-Universität Erlangen-Nürnberg, Nägelsbachstraße 25, 91052 Erlangen, Germany

Department of Physics, Friedrich-Alexander-Universiät Erlangen-Nürnberg, Erwin-Rommelstraße 1, 91058 Erlangen, Germany

**1 Experimental Details**

Raman Spectroscopy: Raman spectroscopic characterization was carried out on a LabRAM Aramis confocal Raman microscope (Horiba) with a laser spot size of about 1 µm (Olympus LMPlanFl 50x LWD, NA 0.50) in backscattering geometry. As excitation source a green laser with $\lambda_{exc}$ = 532 nm was used and the incident laser power was kept as low as possible (1.35 mW) to avoid any structural sample damage. Spectra were recorded with a CCD array at -70 °C – grating: 600 grooves/mm. Exact sample movement was provided by an automated xy-scanning table. Calibration in frequency was carried out with a HOPG crystal as reference.

Glove Box: Sample synthesis and preparation was carried out in an Ar-filled LABmasterpro sp glove box (MBraun), equipped with a gas purifier and solvent vapor removal unit: oxygen content < 0.1 ppm, water content < 0.1 ppm.

Chemicals and graphite were purchased from Sigma Aldrich. The 1,2-dimethoxythane (DME) was distilled twice and stored under Ar over molecular sieves 4 Å for 3 days. The dry solvent ($H_2O$ < 10 ppm) was pump freezed 5 times to receive oxygen free DME.

Procedure used for the preparation of arylated monolayer/bilayer graphene at RT:

In a glove box (Ar-filled, 0.1 ppm $H_2O$ and 0.1 ppm $O_2$) a liquid $NaK_3$ alloy was prepared by melting 3.9 mmol Na and 7.7 mmol K at RT. The alloy was then added to 10 mL of fresh distilled and degassed DME and stirred 1 h to give a deep blue solution. One drop of this solution was dropped on a $Si/SiO_2$ wafer with micromechanical cleaved graphene. Afterwards, a drop of *bis*-(4-(*tert*-butyl)phenyl)iodonium hexafluorophosphate in DME was added. The wafer was washed with DME after 10 min reaction time and removed from the glove box. Outside the glove box the wafer was washed twice with 2-propanol and once with acetone to remove salt residues. Finally, it was dried at 40 °C for 1 h.

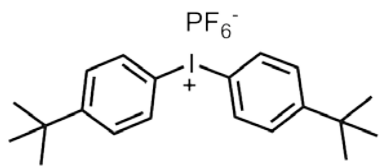

**Figure S1**. *bis*-(4-(*tert*-butyl)phenyl)iodonium hexafluorophosphate.

Procedure used for the *in situ* measurement:

In a glove box (Ar-filled, 0.1 ppm $H_2O$ and 0.1 ppm $O_2$) a liquid $NaK_3$ alloy was prepared by melting 3.9 mmol Na and 7.7 mmol K at RT. The alloy was then added to 10 mL of fresh distilled and degassed 1,2-DME and stirred 1 h to give a deep blue solution At RT one drop of this solution was dropped on a $SiO_2/Si$ wafer with micromechanical cleaved graphene. The wafer was transferred to an *in situ* measurement cell, which was filled with Ar and the activated graphene was removed from the glove box.

Procedure for functionalization without activation:

A drop of *bis*-(4-(*tert*-butyl)phenyl)iodonium hexafluorophosphate in DME was added to a mechanical exfoliated graphene on a $Si/SiO_2$ wafer at RT. After 10 min the wafer was washed with 2-propanol, water and acetone. Finally, the wafer was dried at 40 °C for 1 h.

## 2 Additional Analytical Data

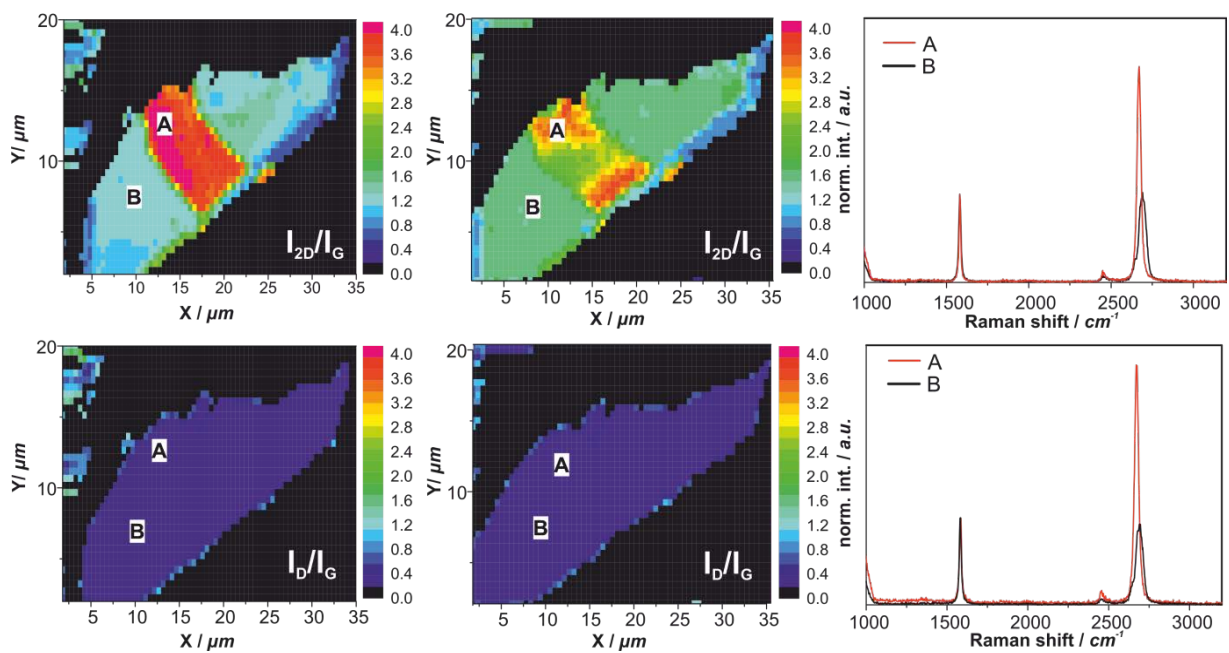

**Figure S2.** Scanning Raman Microscopy of mechanically exfoliated graphene: (left top and bottom) $I_{2D}/I_G$ and $I_D/I_G$ ratio map of pristine flake, (middle top and bottom) $I_{2D}/I_G$ and $I_D/I_G$ ratio map of flake after unactivated functionalization and (right top and bottom) exemplary Raman spectra of the monolayer area A and the bilayer area B pristine (top) and flake after unactivated functionalization (bottom).

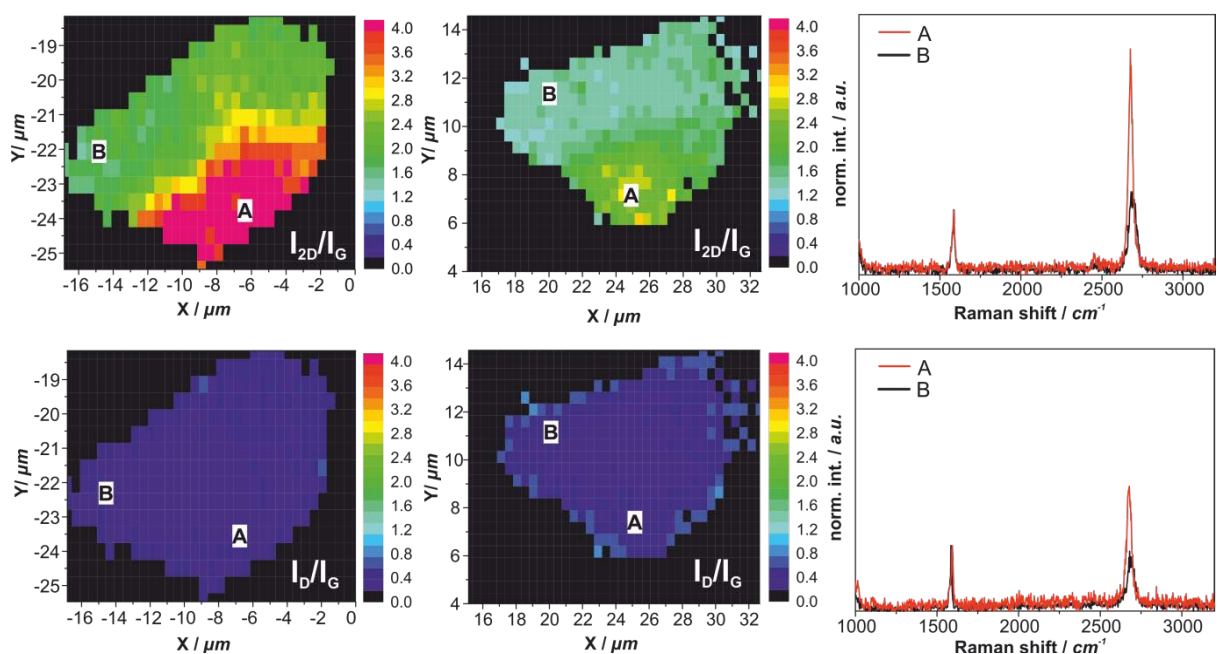

**Figure S3.** Scanning Raman microscopy images of mechanically exfoliated graphene: (left top and bottom) $I_{2D}/I_G$ and $I_D/I_G$ ratio map of pristine flake, (middle top and bottom) $I_{2D}/I_G$ and $I_D/I_G$ ratio map of activated flake and (right top and bottom) exemplary Raman spectra of the monolayer area A and the bilayer area B pristine (top) and activated flake (bottom).

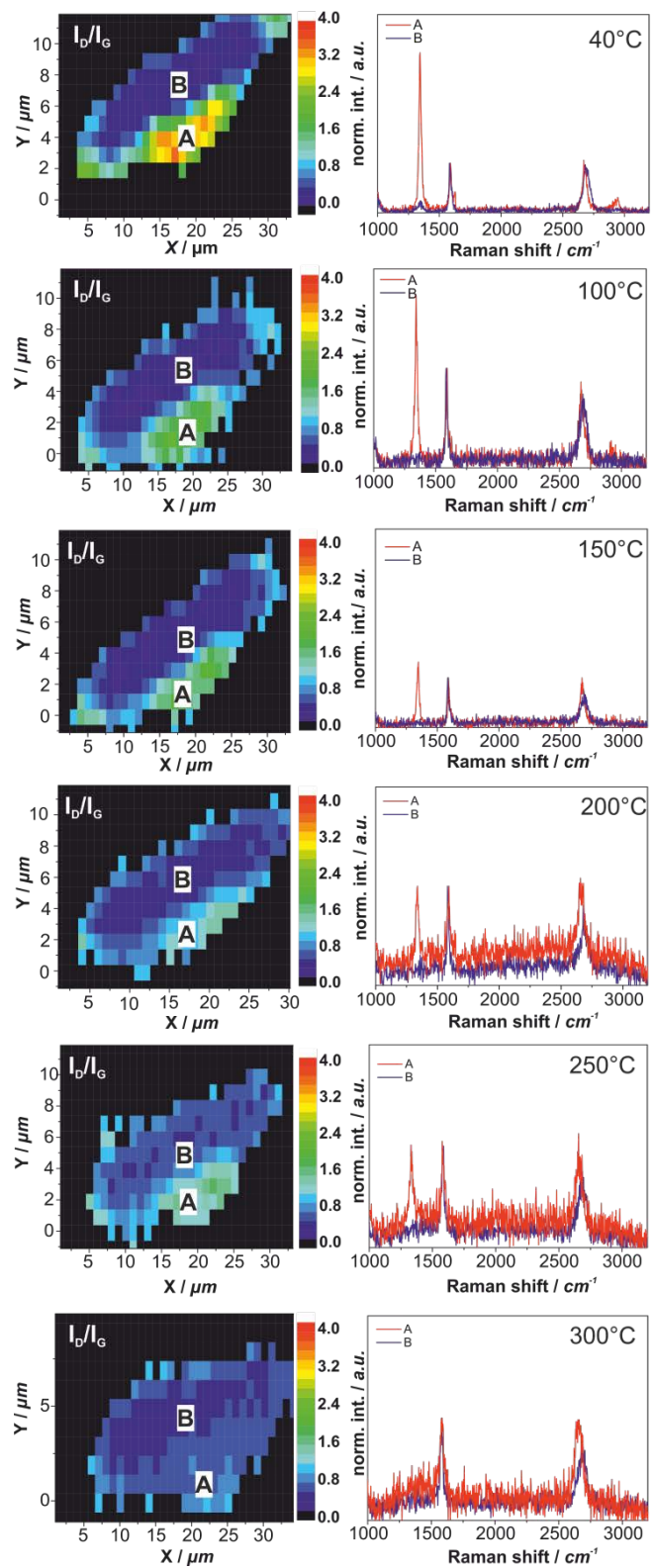

**Figure S4.** Temperature-dependent Raman microscopy images: $I_D/I_G$ ratio map of functionalized graphene flake at 40 °C, 100 °C, 150 °C, 200 °C, 250 °C and 300 °C with point spectra of area A and B.

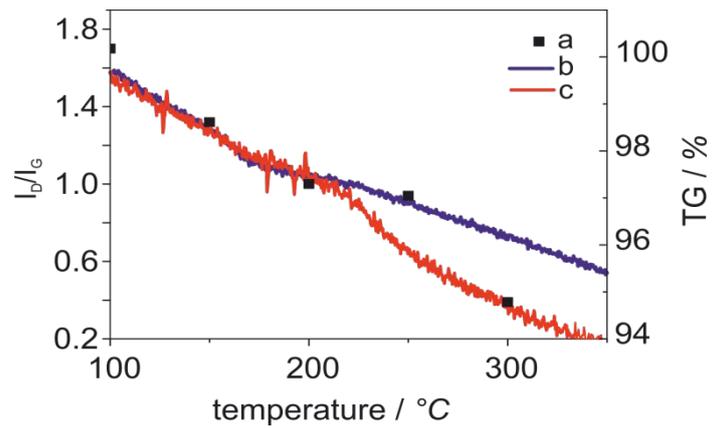

**Figure S5.** Comparison of temperature-dependent Raman spectroscopy $I_D/I_G$ ratios of functionalized monolayer graphene (black squares) and TG profile of functionalized a bulk graphene/graphite (red: $KC_4$ and blue: $KC_8$).

# 3 DFT Calculations

Density-functional theory (DFT) calculations were carried out with the plane-wave code PWscf of the Quantum Espresso software package,[1] using the Perdew-Burke-Ernzerhof PBE exchange-correlation functional,[2] Vanderbilt ultrasoft pseudopotentials,[3] and a plane wave kinetic energy cutoff of 30 Ry. Van der Waals contributions to the total energy and forces were included by Grimme's D2 dispersion correction scheme.[4] A dispersion contribution term was added for all atom pairs within a cutoff distance of 100 Å, except for the Ni-Ni pairs within the substrate, for which the dispersion correction was turned-off. Furthermore, for the dispersion part periodic boundary conditions were only applied parallel to the substrate and graphene layers, but not to the periodic images in the perpendicular direction. $k$-point meshes for Brillouin zone integrations were generated by the Monkhorst-Pack scheme.[5] The density of the k-point meshes with respect to a primitive unit cell was at least (24,24,1) and (18,18,1) for the free-standing and the Ni-supported graphene layers, respectively. The occupation numbers of the electronic states were determined by a Gaussian smearing with a smearing parameter of 0.136 eV. Geometries were optimized by minimizing the atomic forces, with a convergence threshold for the largest residual force of 5 meV/Å.

For bulk Ni we obtained a lattice constant of 3.521 Å with our DFT/PBE setup (using spinpolarized calculations for the ferromagnetic ground state), which deviated only by +0.2% from the experimental value at 0 K of 3.516 Å. The Ni(111) substrate was modeled by periodically repeated slabs with a thickness of 3 atomic layers. The theoretical bulk lattice constant was used for the lateral extension of the unit cells and the repeated images of the slabs were separated by a vacuum region of about 10 Å. In the geometry optimizations the bottom layer of the slab was kept fixed and only the upper two layers were allowed to relax. Graphene and Bernal-stacked bilayer graphene layers were added at on-top positions of the Ni substrate in order to allow the formation of chemical bonds between C and Ni atoms. The degree of graphene functionalization by methyl groups was changed by using different unit cell sizes. Specifically, we used (2 × 2), (√3 × √3) and (2 × 1) cells with one methyl group and a (2 × 1) cell with two methyl groups for representing graphene functionalizations of 25%, 33%, 50% and 100%, respectively.

Binding energies (representing the strength of the C-C bond of a methyl group bound to a graphene layer) were calculated by taking the energy difference between the sum of the energy of the structure without methyl group and the methyl radical and the energy of the final structure with the methyl group attached to the mono- or bilayer graphene layers. The atomic configurations and the binding energies for methyl groups on free-standing graphene and Ni-supported mono- and bilayer graphene in the limit of high surface functionalization are summarized in Figure S6.and Figure S7, respectively.

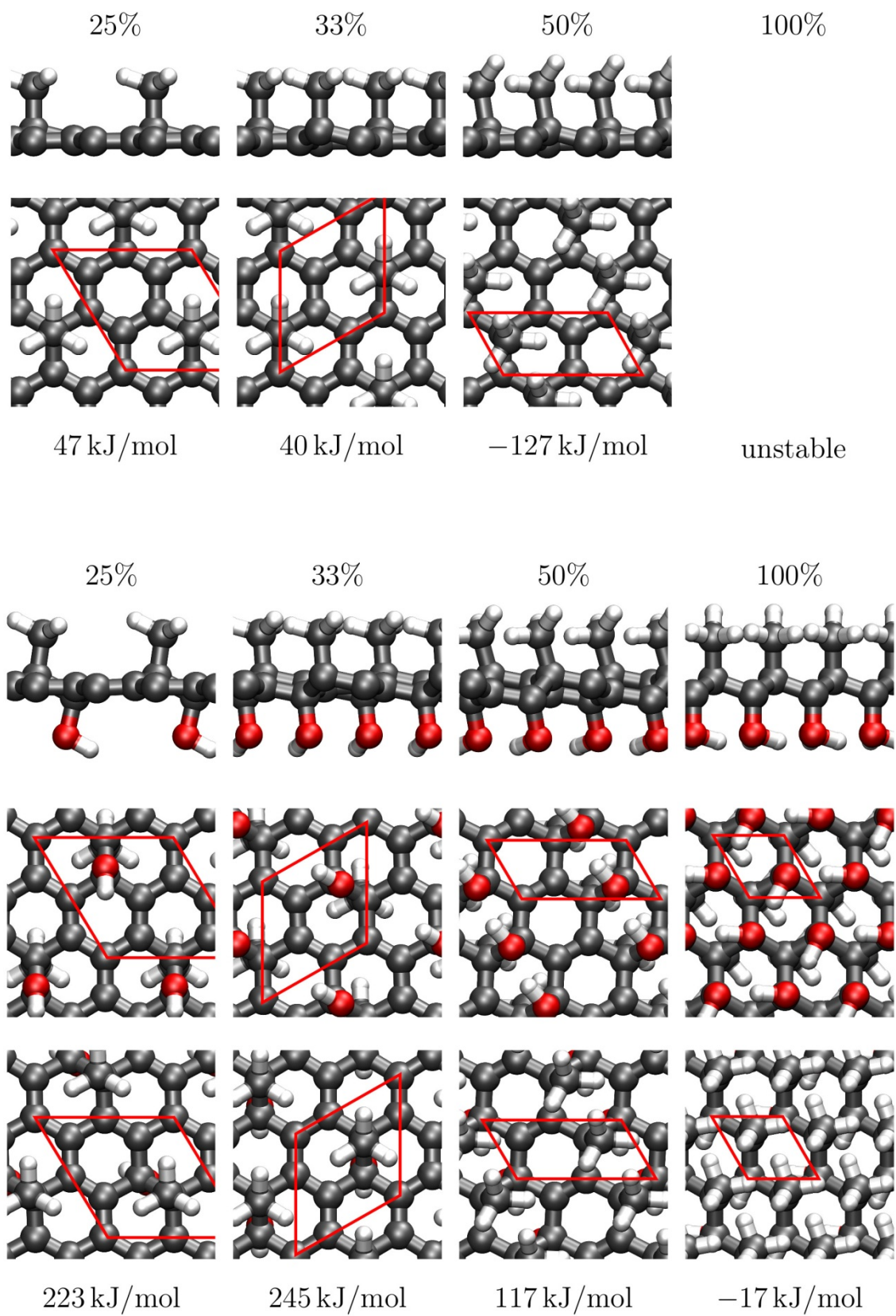

**Figure S6.** Binding energy and atomic structure of methyl groups on single layer graphene without (top) and with saturation of the opposite side by OH groups (bottom) depending on the degree of surface functionalization. C atoms are shown in black, H atoms in white and O atoms in red. The unit cell is indicated by red lines.

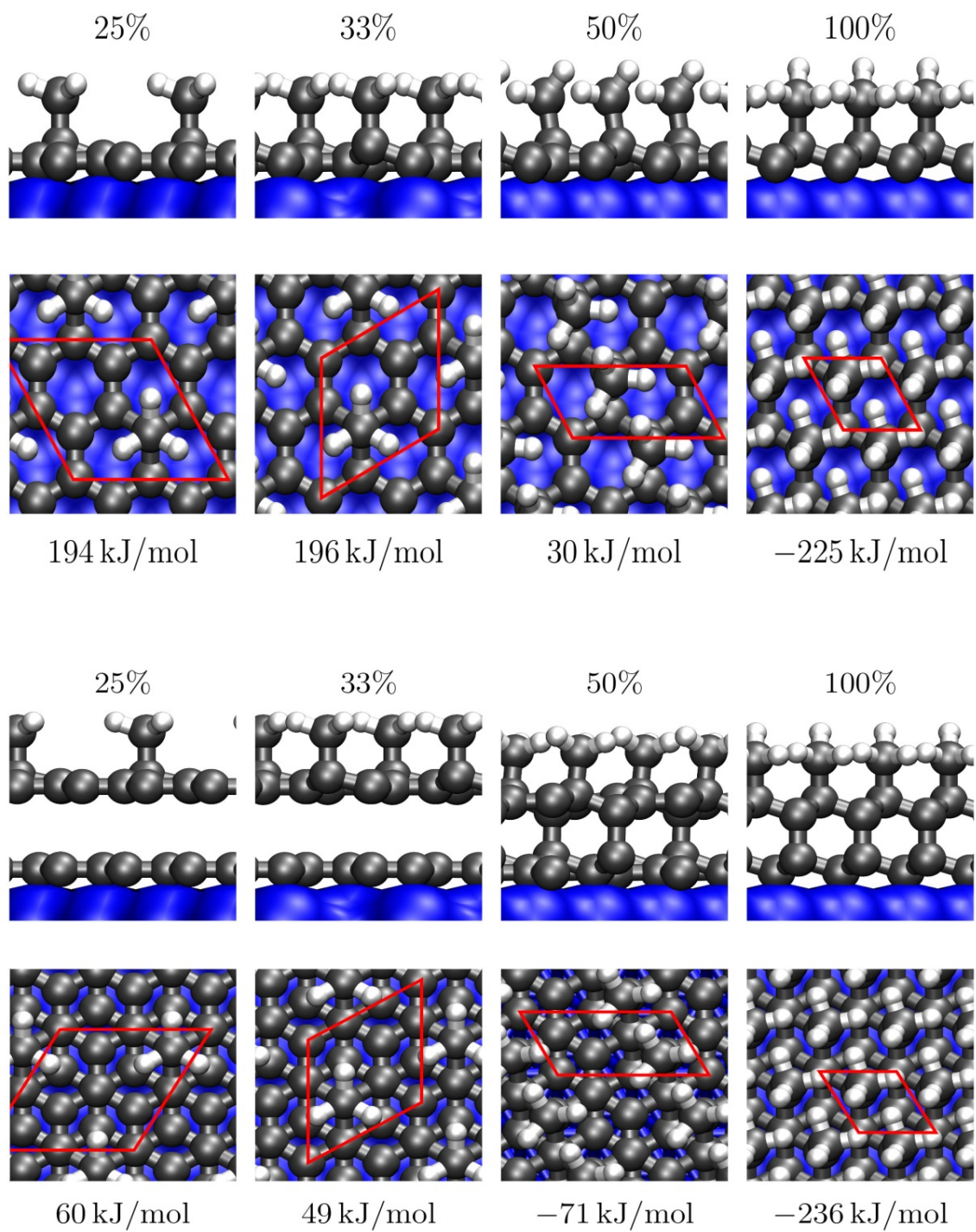

**Figure S7.** Binding energy and atomic structure of methyl groups on single layer (top) and bilayer (bottom) graphene supported on a Ni(111) substrate depending on the degree of surface functionalization. C atoms are shown in black, H atoms in white and the Ni substrate in blue. The unit cell is indicated by red lines.

# 4 References

(1) Giannozzi, P.; Baroni, S.; Bonini, N.; Calandra, M.; Car, R.; Cavazzoni, C.; Ceresoli, D.; Chiarotti, G. L.; Cococcioni, M.; Dabo, I.; Dal Corso, A.; de Gironcoli, S.; Fabris, S.; Fratesi, G.; Gebauer, R.; Gerstmann, U.; Gougoussis, C.; Kokalj, A.; Lazzeri, M.; Martin-Samos, L.; Marzari, N.; Mauri, F.; Mazzarello, R.; Paolini, S.; Pasquarello, A.; Paulatto, L.; Sbraccia, C.; Scandolo, S.; Sclauzero, G.; Seitsonen, A. P.; Smogunov, A.; Umari, P.; Wentzcovitch, R.M. QUANTUM ESPRESSO: A modular and open-source software project for quantum simulations of materials. *J. Phys.: Condens. Matter* **2009**, *21*, 395502.

(2) Perdew, J.P.; Burke, K.; Ernzerhof, M. Generalized Gradient Approximation Made Simple. *Phys. Rev. Lett.* **1996**, *77*, 3865-3868; Erratum: *Phys. Rev. Lett.* **1997**, *78*, 1396.

(3) Vanderbilt, D. Soft self-consistent pseudopotentials in a generalized eigenvalue formalism. *Phys. Rev. B* **1990**, *41*, 7892-7895.

(4) Grimme, S. Semiempirical GGA-Type Density Functional Constructed with a Long-Range Dispersion Correction. *J. Comput. Chem.* **2006**, *27*, 1787-1799.

(5) Monkhorst, H.J.; Pack, J.D. Special points for Brillouin-zone integrations. *Phys. Rev. B* **1976**, *13*, 5188-5192.